\documentstyle[graphicx,aps,preprint,prb,floats,amsmath,amssymb,latexsym]{revtex}
\tightenlines

\newcommand{\CaMinus}{\mbox{Ca$_{1-\delta}$B$_{6}$}}
\newcommand{\CaPlus}{\mbox{CaB$_{6}$}}

\begin{document}
\input epsf

\title{Doping dependence of the electrical and 
thermal transport properties of \CaPlus}
\author{K.~Giann\`o, A.~V. Sologubenko, and H.~R.~Ott}
\address{Laboratorium f\"ur Festk\"orperphysik, ETH 
H\"onggerberg, CH-8093 Z\"urich, Switzerland}
\author{A.~D.~Bianchi and Z.~Fisk}
\address{National High Magnetic Field Laboratory, Florida State 
University, 1800 East Paul Dirac Drive, Tallahassee, Florida 32306, USA}

\maketitle

\begin{abstract}
        The thermoelectric power $S$ and the 
	thermal conductivity $\kappa$ of stoichiometric \CaPlus\, and vacancy-doped 
        \CaMinus\, have been measured between 5 and 300~K. 
	The thermopower of both materials is surprisingly large at room 
	temperature. Across the whole covered temperature range, $S$ is negative 
	and the temperature dependence is most likely dictated by 
	band-structure effects. 
	The empirical interpretation of our data involves a calculation of the
	thermoelectric power by using the Boltzmann equation in the relaxation-time 
	approximation and the assumption of a defect band in the proximity of the lower 
	edge of the conduction band.
	Good agreement with our data is found by considering 
	acoustic-phonon and ionized-impurity scattering for the electrons in 
	the conduction band which is well separated from the valence band.    
	The thermal conductivity $\kappa(T)$, predominantly due to phonons,
	may also be described quite well across the whole covered temperature 
	regime on the basis of a Debye-type relaxation-time 
	approximation and assuming the concurring influence of boundary-, dislocation-,  
	Rayleigh-, and resonant-scattering on the mean free path of the phonons.
\end{abstract}


\section{Introduction}
\label{sec:introduction}

Recent experiments on hexaborides with divalent metal cations of the 
alkaline-earth series, namely Ca, Sr and Ba, have revealed some unusual physical
properties of these materials. 
In particular, an itinerant type of 
weak ferromagnetic order, stable up to temperatures between 600 and 
900~K, has been  observed in \CaPlus\, and related alloys in a very 
narrow range of electron doping.~\cite{Young,Vonlanthen2000}
The electronic properties of alkaline-earth hexaborides place these 
materials close to a metal-insulator 
transition.~\cite{Vonlanthen2000,OttZPhys} 
This has been confirmed theoretically by recent calculations of the electronic 
structure of divalent hexaborides which have shown that, with the 
exception of a small region around the X point of the cubic Brillouin 
zone, the valence- and the conduction bands are separated by a large 
gap of several eV.~\cite{Massidda} 
The slight band overlap is reduced 
if the inter-octahedron B-B bond distances increase, finally 
resulting in the opening of a gap over the entire momentum space. 
It has also been predicted that the transport 
properties of SrB$_{6}$ may be strongly dependent on doping, with 
changes from a $p$- to an $n$-type metal induced by slight shifts of the Fermi 
energy around the zero doping value.~\cite{Massidda}
These observations have led to the speculation that the ground state of 
undoped divalent hexaborides may be characterized by a Bose condensate of 
bound electron-hole pairs, or excitons.~\cite{ZhitomirskyNature,Balents,Barzykin}
Weak ferromagnetism may then develop because of a spontaneous 
time-reversal symmetry breaking via doping.
More recent calculations of the single-particle 
excitation spectrum of \CaPlus, however, have led to the claim that this 
material is not a semimetal but a semiconductor with a minimum band gap of 
0.8~eV.~\cite{Tromp} 
If this possibility is considered, the experimental observations, 
which suggest that binary hexaborides are close to a metal-insulator 
transition, would indicate the presence of a defect band of itinerant charge
carriers. 
In an attempt to resolve the uncertainties with respect to the correct 
description of this material, we have studied the influence of the 
chemical composition on the physical properties of CaB$_{6}$. 
The electrical resistivity, magnetoresistance, low-temperature 
specific heat and the optical conductivity of stoichiometric and 
doped \CaPlus\, samples have been studied and the results have 
been presented in a previous publication.~\cite{Vonlanthen2000}
Below, we present the results of measurements of the thermal conductivity
and the thermopower. 
While data on the electrical conductivity and the electronic 
contribution to the thermal conductivity provide information on the 
density of electronic states $N(E)$ close to the Fermi energy $E_{\rm 
F}$, the thermopower may serve to establish the energy derivative 
of $N(E)$, i.e., an additional important detail of the electronic 
excitation spectrum. 
One of the samples was prepared in
such a way as to obtain material with a close to stoichiometric 
composition, denoted as \CaPlus. The second sample, which we denote as  
\CaMinus, contained a small number of vacancies on the 
calcium sites, inadvertently introduced during the flux-growth procedure 
and leading to a certain degree of self doping which is difficult to 
control. 
This paper is organized as follows.
After a brief description of the sample preparation and the experimental 
methods used in this investigation in section~\ref{sec:experimental}, we present, 
in section~\ref{sec:results}, the results of our measurements and their analysis.
The conclusions are presented in section~\ref{sec:summary}.

\section{Samples and experimental methods}
\label{sec:experimental}

Binary hexaborides can be synthesized in a narrow range of composition, with
the tendency to be boron-rich with concentrations of metal vacancies
up to several percent. Stoichiometric \CaPlus\, crystals can be 
obtained close to the border of the metal-rich phase 
boundary.\cite{OttZPhys,Spear1977} 
For our experiments the single crystals were grown by a slow-cooling 
procedure in aluminum flux~\cite{Fisk1989} starting with a nominal
ratio CaB$_{3}$ for the stoichiometric
\CaPlus\, sample, and with CaB$_{12}$ for the metal-deficient \CaMinus\, sample.  
The crystals were removed from the flux by leaching in a
concentrated sodium hydroxide solution. Subsequent etching with 
HNO$_{3}$ was intended to remove possible surface contaminations.
The samples on which our transport measurements were made were of 
prism-type shape with approximate overall dimensions of 
$4.2\times 0.5\times 0.45\;{\rm mm}^{3}$ for \CaPlus\, and 
$4.7\times 0.45\times 0.4\;{\rm mm}^{3}$ for \CaMinus.  

The thermoelectric power $S$ and the thermal conductivity 
$\kappa$ of both samples were measured simultaneously by means 
of a standard steady-state heat-flow technique. 
A commercial $^{4}$He gas-flow cryostat was used for cooling the
sample holder. At one end of a prism-shaped sample, the thermal
contact to a copper heat sink was achieved by using
high-conductance silver epoxy. The sample heater, consisting of a
$100\;\Omega$ ruthenium-oxide chip resistor, was attached to the
other end of the prism by the same method.
Joule heating 
caused by heater currents of the order of a few mA provided the necessary 
heat-flow and hence a thermal gradient along the crystals. 
In order to measure both the temperature difference $\Delta T$ 
between two contacts mounted perpendicularly to the heat flow
as well as the thermoelectric voltage 
of our samples with respect to chromel, we used two pairs of 
calibrated 0.025 mm \mbox{Au-Fe} (0.07 at. \%) versus chromel 
thermocouples.~\cite{RevSciInstrBougrine} 
In order to minimize spurious thermal voltages, the thermocouple leads 
were connected to uninterrupted copper wires 
reaching three home-built low-noise voltage amplifiers, mounted directly
on top of the cryostat insert.  

\section{Experimental results and Analysis}
\label{sec:results}

\subsection{Thermoelectric power}\label{sec:TEP}

\begin{figure}
        \begin{center}
	\leavevmode
	\epsfxsize=0.7\columnwidth \epsfbox {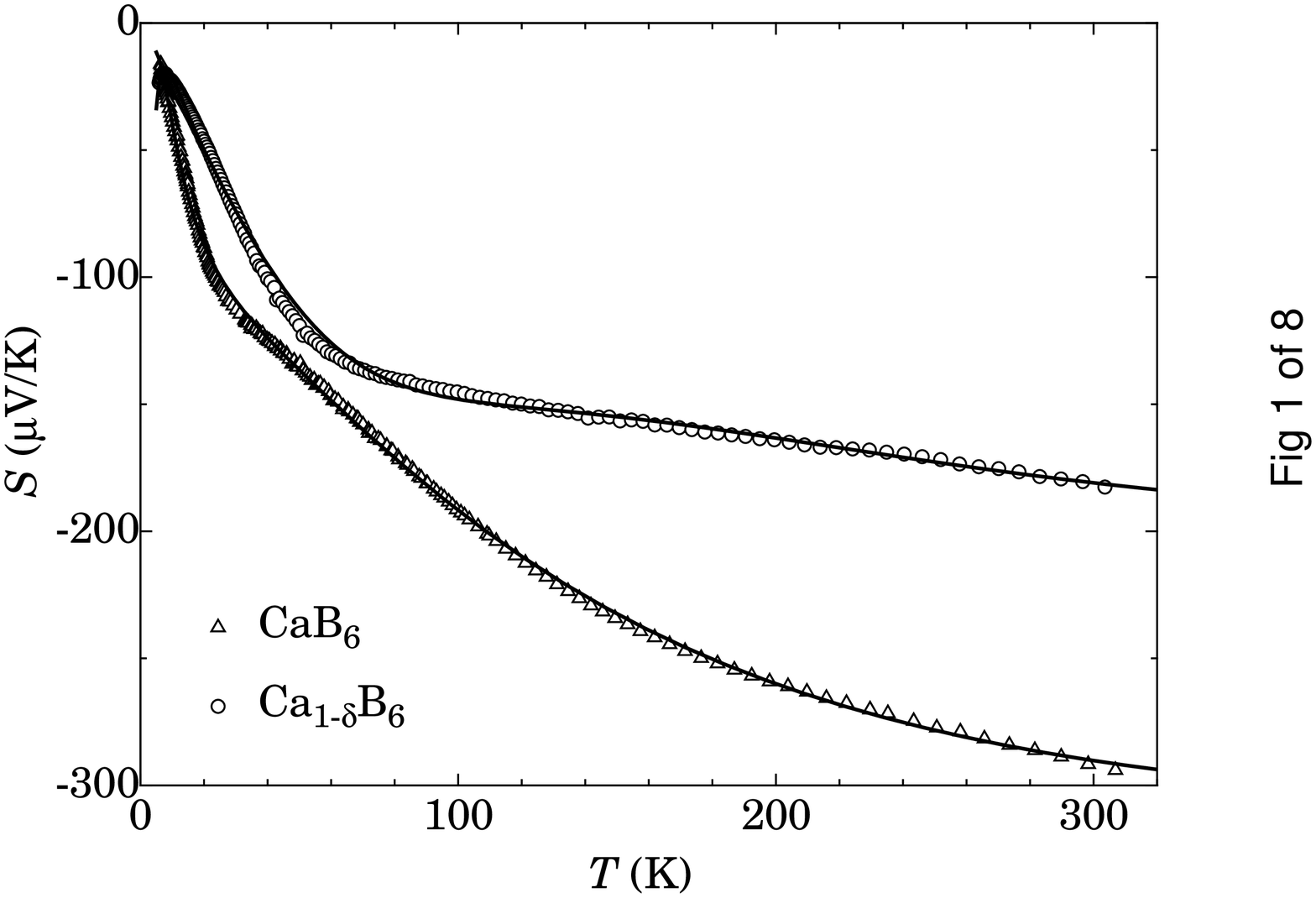}
	\vspace{1.0ex plus 0.5ex minus 0.2ex} 
	\caption{
	    	Temperature dependence of the thermoelectric power 
           	$S$ of \CaPlus\, and \CaMinus\, on linear scales. The 
           	solid lines represent fits to our data, as explained in 
           	the text.
	}
           \protect\label{CaB6S}
	\end{center}
\end{figure}

The thermoelectric power $S(T)$ of both samples is shown in 
Fig.~\ref{CaB6S} on linear scales. 
The negative sign of $S$ clearly demonstrates the dominant $n$-type character of the 
investigated crystals. 
The rather large values of $S$, if compared to the typical values 
observed for the thermoelectric power of common metals, are an
indication for the low itinerant charge-carrier concentration 
in these materials, compatible with the very low values of the electrical 
conductivity reported previously.~\cite{Vonlanthen2000} 
The overall temperature dependence of $S$ is obviously non-linear, with features 
that are reminiscent of band-structure effects affecting the thermopower
of elements and crystalline alloys.~\cite{Barnard}
Nevertheless, below $T^{*} = 20$ and 40~K for \CaPlus\, and \CaMinus, 
respectively, the curves show an 
approximately linear temperature dependence.
A negative thermoelectric power varying linearly with $T$ is usually identified
as the diffusion thermopower $S$ of metals, where the 
Fermi-energy $E_{\rm F}$ is 
located within the conduction band. In the free-electron approximation, 
\begin{equation}\label{EFS}
    S = -\frac{\pi^{2}k_{\rm B}^{2}T}{3|e|E_{\rm F}/s},
\end{equation}
with $E_{\rm F}$ as the Fermi-energy measured from the bottom of the 
conduction band, and $s$ a factor, typically of the order of unity, 
describing the energy dependence of the scattering time
$\tau\propto E^{s-3/2}$ for the scattering mechanism dominating
the low temperature behaviour.~\cite{Barnard}
From the average slope of $S(T)$ below $T^{*}$, we may thus calculate 
a rescaled Fermi energy $E_{\rm F}/s$.
These values are $E_{\rm F}/s= 4.5$ and 8.7~meV 
for \CaPlus\, and \CaMinus, respectively.
Considering the entire temperature dependence of $S$, 
in particular the abrupt changes of slopes of $S(T)$ around $T^{*}$, 
suggests that ultimately, $S(T)$ cannot be explained by simply taking into
account electronic states in the conduction band alone. 
In the following, we try to identify the possible reasons for the measured departure
from linearity in the temperature dependence of $S$.

With the common assumption that the current densities ${\bf j}$ and ${\bf w}$ for 
electrical and thermal transport, respectively, respond linearly to their driving 
forces, we can write
\begin{align}\label{jw}
    \begin{split}
	{\bf j}  &= 
	e^{2}L_{0}{\bf E} - \frac{e}{T}L_{1}{\bf \nabla_{\rm r} 
	T}, \\
	{\bf w}  &= 
	eL_{1}{\bf E} - \frac{1}{T}L_{2}{\bf \nabla_{\rm r} T}, \\
    \end{split}
\end{align}
where $\bf E$ is the external electric field and ${\bf \nabla_{\rm r} T}$ is the 
temperature gradient.  
In Eq.~\ref{jw}, the transport coefficients $L_{\rm i}$ (i=0,1,2) explicitly fulfill  
the Onsager relations.~\cite{Blatt}
For ${\bf j} = 0$, we obtain ${\bf w} = - \kappa_{\rm el}{\bf\nabla_{\rm r}}T$ 
and ${\bf E} = S {\bf \nabla_{\rm r}}T$, whereas for ${\bf\nabla_{\rm r}}T = 0$, 
we get ${\bf j} = \sigma {\bf E}$. 
These are the equations defining the electronic contribution to 
the thermal conductivity $\kappa_{\rm el}$, the thermoelectric power $S$, 
and the electrical conductivity $\sigma$, such that

\begin{subequations}\label{kappaSsigma}
\begin{align}
 \kappa_{\rm el} &= 
 \frac{1}{T}\left(L_{2}-\frac{L_{1}^{2}}{L_{0}}\right),\label{kappa}\\
 S &= -\frac{1}{|e|T}\frac{L_{1}}{L_{0}},\label{Sdiff}\\
 \sigma &= e^{2}L_{0}.\label{sigma}
\end{align}
\end{subequations}
The integrals $L_{\rm i}$ (i=0,1,2), are defined as
\begin{equation}\label{Lsi}
    L_{\rm i} = -\int_{-\infty}^{+\infty} \tilde{\sigma}(E)(E-E_{\rm 
    F})^{i}\frac{\partial f_{0}}{\partial E}dE,
\end{equation}
with $\tilde\sigma(E)$ representing the conductivity spectrum, including 
all system-dependent features. 
By partial integration of Eq.~\ref{Lsi} and by using the Sommerfeld 
expansion scheme, we get 
\begin{equation}\label{dSigmaS}
    S = -\frac{\pi^{2}k_{\rm B}^{2}T}{3|e|}\left(\frac{\partial \ln 
    \tilde\sigma(E)}{{\partial E}}\right)_{E=E_{\rm F}},
\end{equation}
which approximates Eq.~\ref{Sdiff} very well at temperatures 
$k_{\rm B}T \ll E_{\rm F}$. Since we intend to interpret our
data over the entire covered temperature range, we will not use the low-temperature 
approximation represented by Eq.~\ref{dSigmaS}, in our analysis. 
For our purposes the general form of $S$ shown in Eq.~\ref{Sdiff} is 
more suitable.  

Since $\tilde\sigma(E)>0$, the coefficient $L_{0}$ is strictly 
positive and the sign of $S$ is determined by the sign of the integral $L_{1}$.
By inspecting the integrand of $L_{1}$, represented in Eq.~\ref{Lsi}, 
and also by considering Eq.~\ref{Sdiff}, it may be concluded that the states with energies
higher than the chemical potential, i.e., $E>E_{\rm F}$,
provide a negative contribution to the thermoelectric 
power, whereas those states which are located below $E_{\rm F}$ contribute with
the opposite sign.
At low temperatures, where only the states very close to $E_{\rm F}$ 
contribute to the electronic transport, the measured $S$ values are negative.
This leads us to conclude that $\tilde\sigma(E)$ must increase 
monotonously across $E_{\rm F}$, so that the states contributing 
negatively to $S(T)$ in $L_{1}$ acquire a larger weight than those located below 
the Fermi-energy, which contribute positively to $S(T)$. 
In order to reproduce the observed reduction 
of the slope of $S(T)$ at $T^{*}$ in the calculation, however, we are forced to 
introduce a ``feature'' in $\tilde\sigma(E)$ centered at an energy $E^{*}$ such that 
$|E^{*}-E_{\rm F}|\sim k_{\rm B}T^{*}$. Since the absolute value of 
$\partial S/\partial T$ decreases above $T^{*}$, the feature in 
$\tilde\sigma(E)$ must either reduce the 
negative contribution to $S(T)$ given by the states above $E_{\rm F}$, 
or else enhance the positive contribution to $S(T)$ of the 
states below $E_{\rm F}$. 
In what follows, we assume that in our case the second possibility is realized
and we add an additional part to the $\tilde\sigma(E)$ spectrum, 
decreasing linearly with increasing energy below $E_{\rm F}$. The 
complete chosen $\tilde\sigma(E)$ spectrum, corresponding to the 
Eqs.~\ref{sigman} and \ref{sigmap} below, is shown in Fig.~\ref{CaB6Motivation}.

Based on this preliminary analysis and the model outlined above, we have 
calculated the electrical transport properties of \CaPlus\, and \CaMinus.
The Fermi energy is fixed within a parabolic conduction band. Close to the 
bottom of this band we position an additional band, which may 
be interpreted as a defect band.
With these assumptions, the conductivity spectrum may be written as
\begin{equation}
    \tilde\sigma(E) = \tilde\sigma^{n}(E)+\tilde\sigma^{p}(E).
\end{equation}

For an isotropic parabolic conduction band, the Boltzmann equation in 
the relaxation time approximation leads to
\begin{equation}\label{sigman}
  \tilde{\sigma}^{n}(E) = 
  \begin{cases}
      0                                            & (E<E_{\rm n}) \\
      \frac{1}{3\pi^{2}m_{\rm 
      n}}\tau(E)\left(\frac{2m_{\rm n}}{\hbar^{2}}\right)^{3/2}(E-E_{\rm 
      n})^{3/2}     & (E>E_{\rm n}),
   \end{cases}
\end{equation}
where the bottom of the conduction band has been 
fixed to $E_{\rm n} = 0\;{\rm meV}$. 
By setting $\tilde\sigma(E) = \tilde\sigma^{n}(E)$ in Eq.~\ref{dSigmaS}, it 
is easy to recover Eq.~\ref{EFS}. 
For the effective mass of 
the conduction electrons we used $m_{n} = 0.28 m_{0}$,~\cite{Massidda,Tromp} 
with $m_{0}$ as the free electron mass. 
We only considered two scattering mechanisms of 
the charge carriers in the conduction band, i.e., scattering by  
acoustic lattice vibrations, approximated by a rate~\cite{Ziman,Nag}  
\begin{equation}
    \tau_{\rm lv}^{-1} = \alpha E^{1/2}T,
\end{equation}
and scattering by ionized impurities, with a rate~\cite{Nag}
\begin{equation}
    \tau_{\rm ii}^{-1} = \beta E^{-3/2}, 
\end{equation}
where $\alpha$ and $\beta$ are two constants to be determined by the 
fitting procedure. If the defect states which are forming a band below
the bottom of the conduction band coexist with localized states, the 
latter may act, when ionized, as scattering centers for the 
electrons. This justifies considering a term $\tau_{\rm ii}^{-1}$ 
in the scattering rate of the electrons. 
As we will show below, the existence of localized donor states may 
also be compatible with our thermal conductivity data. 
Another possibility is to attribute the rate $\tau^{-1}_{\rm ii}$ to the 
vacancies at the Ca$^{2+}$-ion sites which would act as negatively 
charged scattering centers. 
With the common assumption that the two scattering mechanisms do not 
interfere with each other, we may use Matthiessen's rule in the form 
$\tau = (\tau_{\rm lv}^{-1}+\tau_{\rm ii}^{-1})^{-1}$ for the electrons' 
average scattering time.

For the defect band, we postulate~\cite{Enderby}
\begin{equation}\label{sigmap}
  \tilde{\sigma}^{p}(E) =
  \begin{cases}
      \gamma (E_{\rm me}-E)
      & (E<E_{\rm me}) \\
      0         & (E>E_{\rm me}),
  \end{cases}
\end{equation}
with a mobility edge $E_{\rm me}$ and $\gamma$ a constant.
Since the quantities $L_{0}$ and $L_{1}$ which dictate the 
temperature dependence of the thermopower, appear in the ratio 
$L_{1}/L_{0}$, $S$ depends only on the relative 
magnitude of the constants $\alpha$, $\beta$, and $\gamma$.
This allows some reduction of the number of fitting parameters.
In Fig.~\ref{CaB6S}, the solid curves represent the results of the 
calculation of $S(T)$. 
By considering the simplicity of the model and the small number
of free parameters ($\alpha/\gamma, \beta/\gamma, E_{\rm F}, \text{and } 
E_{\rm me}$), 
the agreement with the experimental data is quite remarkable. 
The Fermi energies which emerge from the fitting procedure amount
to 14 and 28~meV for \CaPlus\, and \CaMinus, respectively, 
corresponding to 0~K charge-carrier densities $n_{\rm c}$ of 
$1.1\times 10^{24}\;{\rm m^{-3}}$ and 
$3.2\times 10^{24}\;{\rm m^{-3}}$, or to $0.8\times 10^{-4}$
and $2.3\times 10^{-4}$ charge carriers per unit cell, respectively. 
Preliminary results of low-temperature Hall effect measurements performed
on a \CaPlus\, sample with the same resistivity as our vacancy-doped 
sample lead, when 
interpreted with a single-band model, 
to a charge carrier density of $4.2\times 10^{24}\;{\rm m^{-3}}$ 
(Ref.~\onlinecite{Waelti}), hence compatible with our results. 
The value of the mobility edge $E_{\rm me}$, introduced to 
reproduce the kink in $S(T)$ at the observed temperatures is approximately $+8$~meV 
for both samples. 
We note, however, that this value strongly depends on the choice of the energy exponent
of the low-temperature scattering term.
In this sense we have to admit that this result depends on the employed model.

\begin{figure}[t]
        \begin{center}
	\leavevmode
	\epsfxsize=0.7\columnwidth \epsfbox  {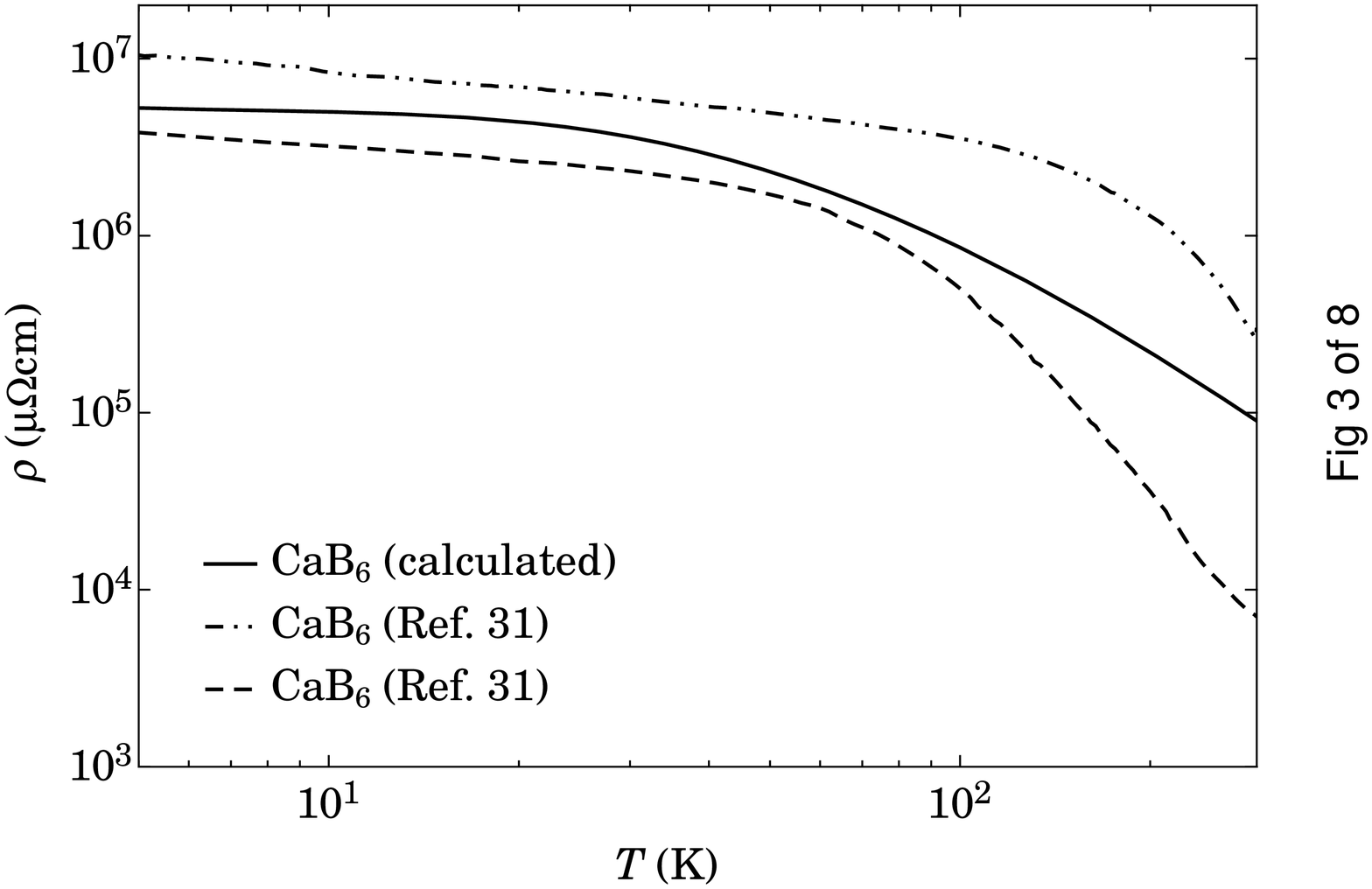}
	\vspace{1.0ex plus 0.5ex minus 0.2ex} 
           \caption{The solid line represents the expected temperature dependence 
	   of the electrical resistivity of \CaPlus\, which is based on the 
           postulated conductivity spectrum $\tilde\sigma(E)$, as 
           explained in the text. Also shown in the figure as dotted 
           and broken line are the measured temperature 
           dependencies of the resistivities of two samples originating from the
	   same batch as the \CaPlus\, sample measured in this 
	   investigation.}
           \protect\label{CaB6rho}
	\end{center}
\end{figure}

Our calculation of the thermoelectric power relies essentially on the
model of the conductivity spectrum (modulo a factor), which we may 
now test by calculating other electronic transport properties such as the 
electrical resistivity $\rho(T)$ and the thermal conductivity $\kappa_{\rm el}(T)$ 
and compare the results with the experimental data. In Fig.~\ref{CaB6rho}, we 
have plotted the rescaled electrical resistivity of \CaPlus\, 
calculated with Eq.~\ref{sigma}, together with the experimental data taken on two 
samples of the same batch. 
The agreement between the calculated curve and the experimental data is 
acceptable, especially if one considers primarily the temperature dependence. 

By using Eqs.~\ref{sigma}, \ref{Lsi}, and \ref{sigman}, combined with 
the measured resistivity curves shown in Fig.~\ref{CaB6rho}, we conclude
that the mean free path $l_{\rm F} = \hbar k_{\rm F}\tau(E_{\rm F})/m_{\rm n}$ 
of the conduction electrons in \CaPlus\, is less than 100~\AA. 
The same length is obtained, obviously enough, by employing the Drude formula
$\sigma = n e^{2} \tau/m_{n}$ and by using $n=n_{\rm c}$, the value 
cited above for the charge carrier concentration of \CaPlus. 
Since the electrons' wave vector at the Fermi-energy is of the order of 
$10^{-2}$~\AA$^{-1}$, it seems fair to assume that the Ioffe's criterion for 
localization ($k_{\rm F}l_{\rm F}<1$) is close to be fulfilled in \CaPlus.

\subsection{Thermal transport}

\begin{figure}[t]
    \begin{center}
	\leavevmode
	\epsfxsize=0.7\columnwidth \epsfbox  {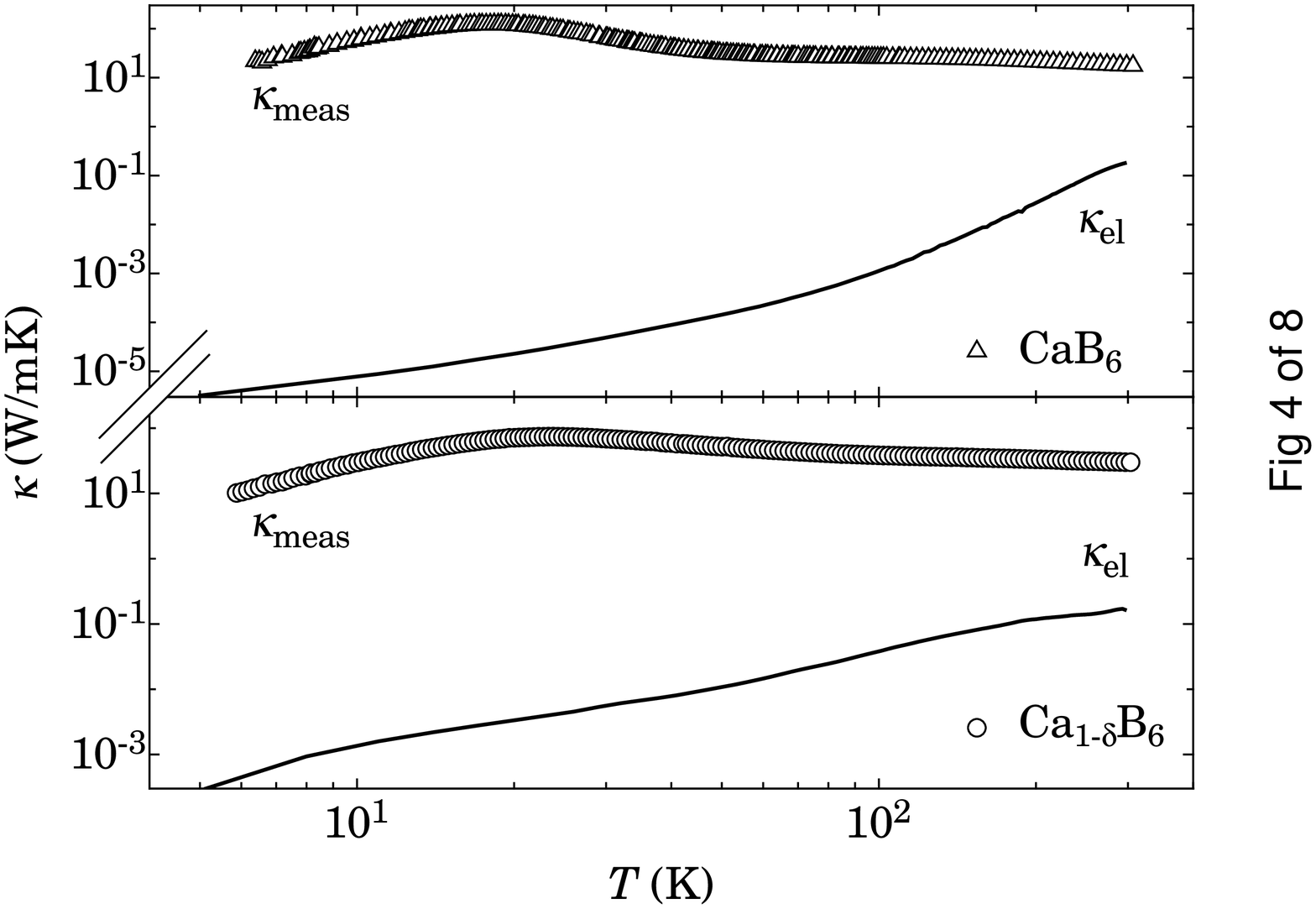}
 	\vspace{1.0ex plus 0.5ex minus 0.2ex} 
            \caption{The measured thermal conductivities $\kappa_{\rm 
            meas}$ of \CaPlus\, and \CaMinus\, and their calculated electronic 
            contributions $\kappa_{\rm el}$ (see text) as a function of temperature 
            on logarithmic scales.}
            \protect\label{CaB6kappaEl}
 	\end{center}
\end{figure}

In Fig.~\ref{CaB6kappaEl}, we show the measured thermal 
conductivities $\kappa_{\rm meas}$ of \CaPlus\, and \CaMinus\, on 
logarithmic scales between 6 and 300~K. 
In the further analysis we assume that only conduction-electron
and lattice contributions to $\kappa_{\rm meas}$ need to be
considered.

\begin{figure}[t]
        \begin{center}
	\leavevmode
	\epsfxsize=0.7\columnwidth \epsfbox  {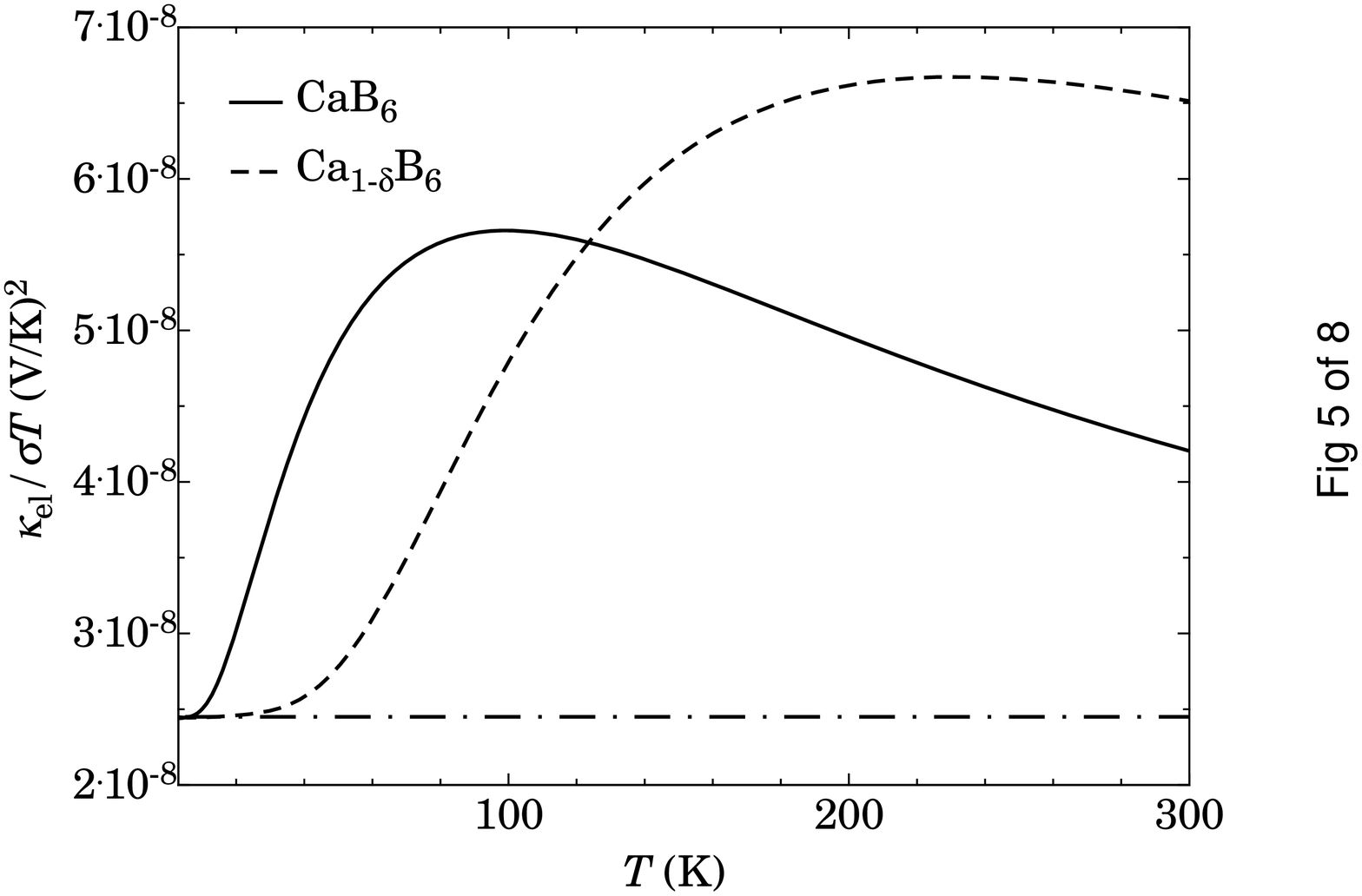}
	\vspace{1.0ex plus 0.5ex minus 0.2ex} 
           \caption{The ratio $\kappa_{\rm el}/\sigma T$ as a 
           function of temperature, calculated by using the 
           Eqs.~\ref{kappa} and \ref{sigma}, with the fit parameters 
           provided by our analysis of $S(T)$ of 
           \CaPlus\, (solid line) and \CaMinus\, (dashed line). The 
           horizontal dashed-dotted line is the Lorenz number 
           $L_{0}=2.45\times 10^{-8}\;{\rm (V/K)^{2}}$.}
           \protect\label{CaB6L0}
	\end{center}
\end{figure}

A measure for the electronic contribution to the thermal conductivity 
is given by the quantity $L^{*}=\kappa_{\rm el}/\sigma T$, which can be calculated 
from the Eqs.~\ref{kappa} and \ref{sigma}. Just like the thermoelectric power $S$, 
$L^{*}$ depends only on the relative magnitude of the constants $\alpha$, 
$\beta$, and $\gamma$. 
In a degenerate one-band model, the value 
\begin{equation}
    \kappa_{\rm el}/\sigma T =  L_{0} = 
    \frac{\pi^{2}}{3}\left(\frac{k_{\rm B}}{e}\right)^{2}
\end{equation}
is denoted as the Lorenz number. The ratio $L^{*}$ can be much higher
in semiconductors than in metals, due to the so-called 
ambipolar diffusion, i.e., the diffusion of electron-hole pairs. 
In Fig.~\ref{CaB6L0} we plotted the temperature dependence of 
$L^{*}$, calculated by using the fitting parameters 
provided by our analysis of $S(T)$, together with the value 
$L_{0}=2.45\times 10^{-8}{\rm (V/K)^{2}}$.  
As expected, $\kappa_{\rm el}/\sigma T$ is equal to the Lorenz number at
very low temperatures. This is so because only the electrons of 
the conduction band contribute to the transport.    
The influence of the ambipolar diffusion to the electronic 
thermal conductivities of \CaPlus\, and \CaMinus\, is manifest 
at intermediate temperatures. Nevertheless, this effect is still quite small if 
compared to enhancements of $L^{*}$ by two orders 
of magnitude observed in intrinsic semiconductors.~\cite{Goldsmid}

The electronic contributions to the thermal conductivities of \CaPlus\, 
and \CaMinus, shown in Fig.~\ref{CaB6kappaEl} as solid lines, 
have beeen calculated by using the Wiedemann-Franz 
relation $\kappa_{\rm el} = L^{*} \sigma T$, with $L^{*}$ calculated from 
the Eqs.~\ref{kappa} and \ref{sigma} as shown in Fig.~\ref{CaB6L0}.
For the electrical conductivity $\sigma$, we used the values which have 
typically been measured on close-to-stoichiometric and vacancy-doped 
samples, respectively.~\cite{Vonlanthen2000,Paschen2000,VonlanthenPC}  
For both hexaborides, the estimated $\kappa_{\rm el}$ is at least two 
orders of magnitude smaller than $\kappa_{\rm meas}$ at all 
temperatures covered in this study. Since the experimental uncertainty 
in our measurements of the thermal conductance is of the order of $0.5\%$, 
the procedure for estimating $\kappa_{\rm el}$ is not critical and may be 
considered as sufficiently accurate for the validity of the following 
considerations.

\begin{figure}[t]
        \begin{center}
	\leavevmode
	\epsfxsize=0.7\columnwidth \epsfbox  {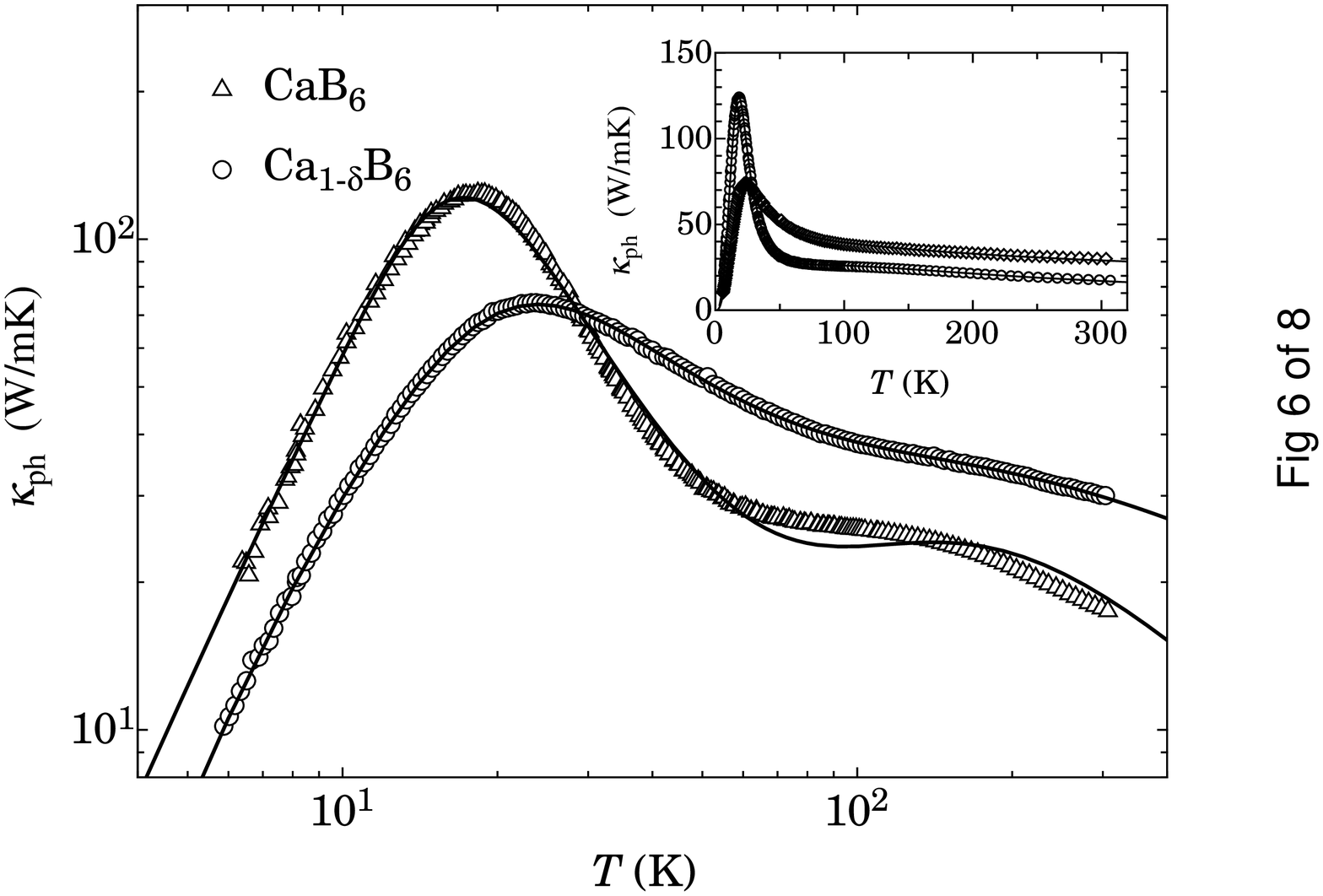}
	\vspace{1.0ex plus 0.5ex minus 0.2ex} 
           \caption{The lattice contribution to the thermal conductivity 
           $\kappa_{\rm ph}$ of \CaPlus\, and \CaMinus\, as a function of 
           temperature on logarithmic scales (on linear scales in the 
           inset). The solid lines 
           correspond to fits to the data as explained in the text.}
           \protect\label{CaB6kappa}
	\end{center}
\end{figure}

The lattice contributions $\kappa_{\rm ph} = \kappa_{\rm meas} - 
\kappa_{\rm el}$ to the thermal conductivities of both samples are shown in 
Fig.~\ref{CaB6kappa}, revealing monotonic increases with $T$ 
up to 18~K and 23~K for \CaPlus\, and \CaMinus, respectively. At 
these temperatures, $\kappa_{\rm ph}$ 
passes over maxima of 124 and 74 $\rm W m^{-1} K^{-1}$, 
respectively, and decreases with varying slopes up to room temperature. 
We note that at low temperatures $\kappa_{\rm ph}$ of \CaMinus\, 
attains lower values than $\kappa_{\rm ph}$ for \CaPlus. This seems 
reasonable because of an expected enhanced scattering of the phonons 
by defects in the Ca-deficient material. 
The crossing of the curves at a temperature close to 30~K is a rather 
surprising experimental observation which indicates a more efficient 
high-frequency scattering mechanism of the phonons in the 
structurally more perfect sample. 

The following quantitative analysis of our data is based on a relaxation-time 
approximation in the context of the Debye model.
The Debye approximation is valid at temperatures which are low 
enough with respect to the Debye temperature $\Theta_{\rm D}$, where 
we may assume that the acoustic branches $\omega(k)$ of the vibrational modes are 
well approximated by a linear $k$ dependence. 
Since the temperature range covered in our experiment 
is considerably lower than the Debye temperature $\Theta_{\rm 
D}=783$~K,~\cite{Vonlanthen2000} this approximation is justified to 
some extent. 
With this simplification, the thermal conductivity may be calculated from

\begin{equation}\label{Debye}
    \kappa_{\rm ph}^{\rm fit} = 3nk_{\rm B} \left(\frac{T}{\Theta_{\rm 
    D}}\right)^{3}v_{\rm ph}^{2}\int_{0}^{\Theta_{\rm 
    D}/T}\frac{x^{4}e^{x}}{(e^{x}-1)^{2}}\tau(x,T) dx,
\end{equation}

where $\tau(x,T)$ is the mean time between collisions of a phonon, $n$ the 
number density of atoms in the crystal, and $x=\hbar\omega/k_{\rm B} T$. 
By limiting the total density of allowed degrees of 
freedom to $3n$, it is possible to calculate the cut-off frequency 
$\omega_{\rm D}$, which is related to the slope of $\omega(k)$. 
This cut-off frequency defines the Debye temperature $\Theta_{\rm D}$ 
of the material which is, in this sense, a measure of the phonon velocity 
and is usually extracted from low-temperature specific-heat 
or elastic-constant measurements.
For the Debye temperature $\Theta_{\rm D}$, we used the above cited 
value which was extracted from specific-heat measurements on stoichiometric 
\CaPlus.~\cite{Vonlanthen2000} 
Since upon introducing a very small number of calcium vacancies, no 
significant changes in the lattice contribution to the specific heat are expected, 
we used the same Debye temperature for \CaPlus\, and \CaMinus. 
This argument is supported by the observation that there are 
no significant differences between the low-temperature specific heats
of \CaPlus\, and Ca$_{1-x}$La$_{x}$B$_{6}$ with x=0.005.~\cite{Vonlanthen2000}
The number density of atoms in \CaPlus\, can be calculated by using 
the lattice constant $a=4.146$~\AA reported in 
Ref.~\onlinecite{Spear1977}. From the relation 
\begin{equation}
    n=1/6\pi^{2}(k_{\rm B}\Theta_{\rm D}/\hbar v_{\rm ph})^{3},
\end{equation}   
we may also calculate the average velocity of the vibrational 
modes for which we obtain $v_{\rm ph} = 5722 \,{\rm m/sec}$. This 
value is compatible with the average sound velocity $v_{\rm ph} = 4200 
\,{\rm m/sec}$, valid for EuB$_{6}$~\cite{VonlanthenEuB6} when scaled 
with the mass-density factor $\sqrt{\rho_{\rm EuB_{6}}/\rho_{\rm CaB_{6}}}$.

For evaluating the integral in Eq.~\ref{Debye}, we approximate the phonon 
relaxation rate by 

\begin{equation}\label{tau}
    \tau^{-1}(x,T) = (\tau^{-1}_{\rm Cas} + \tau^{-1}_{\rm 
    disl} + \tau^{-1}_{\rm Rayl} + \tau^{-1}_{\rm res})(x,T).
\end{equation}

\begin{figure}[t]
        \begin{center}
	\leavevmode
	\epsfxsize=0.7\columnwidth \epsfbox  {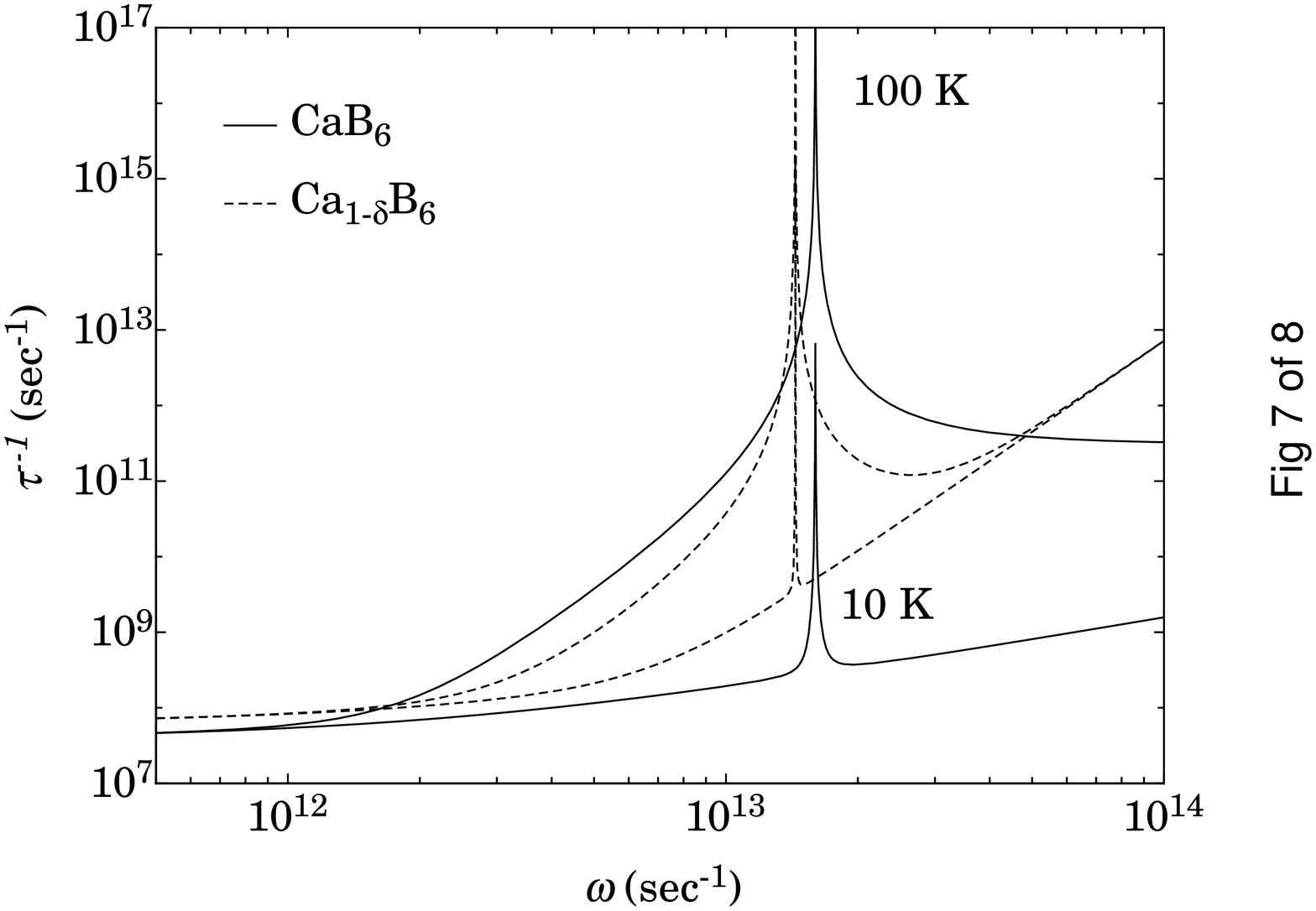}
	\vspace{1.0ex plus 0.5ex minus 0.2ex} 
           \caption{The total phonon relaxation rate $\tau^{-1}$ resulting from 
           our fit (see Eq.~\ref{tau}) as a function of frequency for 
           $T = 10$ and 100~K.} 
           \protect\label{CaB6Rate}
	\end{center}
\end{figure}

The different terms on the right hand side of Eq.~\ref{tau} represent 
the scattering of phonons at grain boundaries, by the strain fields 
surrounding dislocations, 
by point defects and via resonant scattering, respectively. 
The resulting fits are shown in Fig.~\ref{CaB6kappa} as solid lines. 
In what follows, we discuss each term appearing in 
Eq.~\ref{tau} separately. 
The values of the free parameters involved in our fitting 
procedure (see below) are given in Table~\ref{fitpar} and the 
resulting total scattering rates of the phonons are shown as a function 
of frequency in Fig.~\ref{CaB6Rate} for $T=10$ and 100~K.

The temperature and frequency independent scattering length
$l_{\rm Cas}$ in the first term in Eq.~\ref{tau}, $\tau^{-1}_{\rm
Cas} = v_{\rm ph}/l_{\rm Cas}$, is the so-called Casimir lenght,
where the phonon mean free path is limited by the sample
dimensions.~\cite{Casimir} The values of 0.303~mm and 
0.083~mm for \CaPlus\, and \CaMinus, respectively, are of the order of
magnitude of the smallest sample dimensions, resulting from the 
variation of the sample cross section perpendicular to the direction 
of the thermal current.

The second term in Eq.~\ref{tau}, $\tau_{\rm disl}^{-1} = 
A\cdot\omega$, is usually attributed to phonon-electron scattering
but, as pointed out in Ref.~\onlinecite{VonlanthenEuB6}, can also 
be caused by phonon scattering due to strain fields surrounding 
dislocations. 
As may be seen in Table~\ref{fitpar}, this term is slightly larger 
for \CaPlus\, than for \CaMinus.
In view of the higher density of charge carriers in \CaMinus\, 
derived in section~\ref{sec:TEP}, this rules out that mainly
phonon-electron scattering is responsible for this term. 
Phonon-electron scattering may also be excluded by inspecting the 
relevant scattering rate which can be written as~\cite{Ziman}
\begin{equation}
    \tau_{\rm cond}^{-1} = \frac{m_{n}^{2}C_{\rm D}^{2}}{2\pi \rho_{\rm 
    CaB_{6}}v_{\rm ph} \hbar^{3}}\omega,
\end{equation}
with $C_{\rm D}\sim -\frac{2}{3}E_{\rm F}$ as the deformation 
potential. The fit values for $A$ given in Table~\ref{fitpar} result in 
Fermi-energies of the order of 1-2~eV, i.e., values which exceed 
$E_{\rm F}$ evaluated in section~\ref{sec:TEP} by two orders of 
magnitude.
For phonon scattering at dislocations we have~\cite{Klemens} 
\begin{equation}
    \tau_{\rm disl}^{-1} = 0.06 n_{\rm d}(b\gamma_{\rm 
    G})^{2}\omega,
\end{equation}
with $n_{\rm d}$ as the dislocation density, $\gamma_{\rm G}$ 
the Gr\"uneisen parameter, and b the Burger vector. By using 
the lattice constant for $b\sim 4$~\AA and $\gamma_{\rm G}\sim 1$, 
we obtain $n_{\rm d}\sim 2\times 10^{15}\; {\rm m}^{-2}$, a 
plausible result.

The enhanced presence of scattering centers at the vacancy 
sites in \CaMinus\, is unequivocally reflected in the third term of 
Eq.~\ref{tau}. 
Rayleigh scattering by point-like mass defects leads, in the continuum 
approximation, to a scattering rate term 
$\tau_{\rm Rayl}^{-1} = B\cdot\omega^{4}$, with~\cite{Ziman}
\begin{equation}\label{Rayl}
    B = \frac{n_{\rm p}a^{6}}{4\pi v_{\rm ph}^{3}}\left(\frac{\delta 
    M}{M}\right)^{2}.
\end{equation}
In Eq.~\ref{Rayl}, $n_{\rm p}$ is the number density of point defects in 
\CaMinus, $a=4.146$~\AA is its lattice constant, $M$ is 
the mass of a unit cell, and $\delta M$ is the supplementary or 
missing mass due to a point defect in a sphere with the radius of the 
point defect itself. 
Since the extra mass term appears
squared in Eq.~\ref{Rayl}, it is not possible to decide whether
$\delta M$ is positive or negative.  
By assuming that each point defect in \CaMinus\, 
provides one free electron to the ensemble of conduction electrons 
and that for stoichiometric \CaPlus\, $n_{\rm c} = 0$,
we have $n_{\rm p}=3.2\times 10^{24}\;{\rm m^{-3}}$, i.e., the same value as
the electron density calculated in section~\ref{sec:TEP}.  
Again, by using for $B$ the value which we obtained from the fitting procedure, 
we note that $\delta M/M$ is of order unity. 
In view of the rather crude approximations, this result is 
encouraging, as it seems to indicate that the 
concentration of itinerant electrons in \CaPlus\, is equivalent 
to the concentration of impurities (or vacancies) with approximately 
the same mass as the mass of the unit cells. 
The scattering rate term $\tau^{-1}_{\rm Rayl}$ is only of significance 
for fitting the $\kappa_{\rm ph}$ data of the vacancy-doped \CaMinus\, sample.
Our attempts to evaluate the free parameter $B$ for \CaPlus\,  
resulted in a vanishing contribution of $\tau^{-1}_{\rm Rayl}$ to the 
total scattering rate of the stoichiometric hexaboride. Considering 
the estimated error range of a few times $10^{-44}$ sec$^{3}$ for $B$, 
this is at least consistent with a reduced amount of defects in \CaPlus.

The last term appearing in Eq.~\ref{tau} needs a detailed examination. 
Resonant scattering of phonons has already 
been invoked in the analysis of the thermal conductivity data of EuB$_{6}$ 
in a previous investigation.~\cite{VonlanthenEuB6} 
The appearance of $\tau_{\rm res}^{-1}$ in the scattering rate of EuB$_{6}$ 
was intepreted as being due to a resonant scattering of low-momentum 
vibrational states in the dispersionless part of the $\omega(k)$ branches 
with acoustic phonons of constant energy that seem to occupy an
extended region of the Brillouin zone at higher momenta. 
Indeed, a rapid flattening of the LA branches at an energy of approximately 12~meV
has been observed in inelastic neutron scattering experiments performed on 
the trivalent hexaborides XB$_{6}$ (X = Ce, La, Sm).~\cite{Kunii,Smith,Alekseev} 
Because of similar atypical temperature dependencies of $\kappa_{\rm 
ph}$ above 50~K, we follow the arguments given in Ref.~\onlinecite{VonlanthenEuB6} 
and introduce a resonant scattering term of the form
\begin{equation}\label{res}
    \tau_{\rm res}^{-1} = C\cdot\frac{\omega^4}{(\omega_{\rm 
    res}^2-\omega^2)^2}F(T),  
\end{equation}
with
\begin{equation}
    F(T) = \frac{1}{\exp(\hbar\omega_{\rm res}/k_{\rm B}T)-1}
\end{equation}
as the bosonic population factor. 
In our intepretation of the data, the crossing of the thermal 
conductivity curves at \mbox{$T=30$}~K has to be traced back to a much stronger 
resonant scattering rate in \CaPlus. 
As may be seen in Table~\ref{fitpar}, the fit value $C$ is more than one order 
of magnitude larger for \CaPlus\, than for \CaMinus. 

Although, by taking into account the resonant-scattering term, we achieve a 
fairly accurate description of our thermal conductivity data, it seems 
rather unlikely that the large difference in the temperature dependence of
$\kappa_{\rm ph}$ may be accounted for by considering phonon-phonon scattering
processes alone, especially if we consider the only small difference in 
chemical composition between the two materials.

\begin{figure}[t]
        \begin{center}
	\leavevmode
	\epsfxsize=0.7\columnwidth \epsfbox {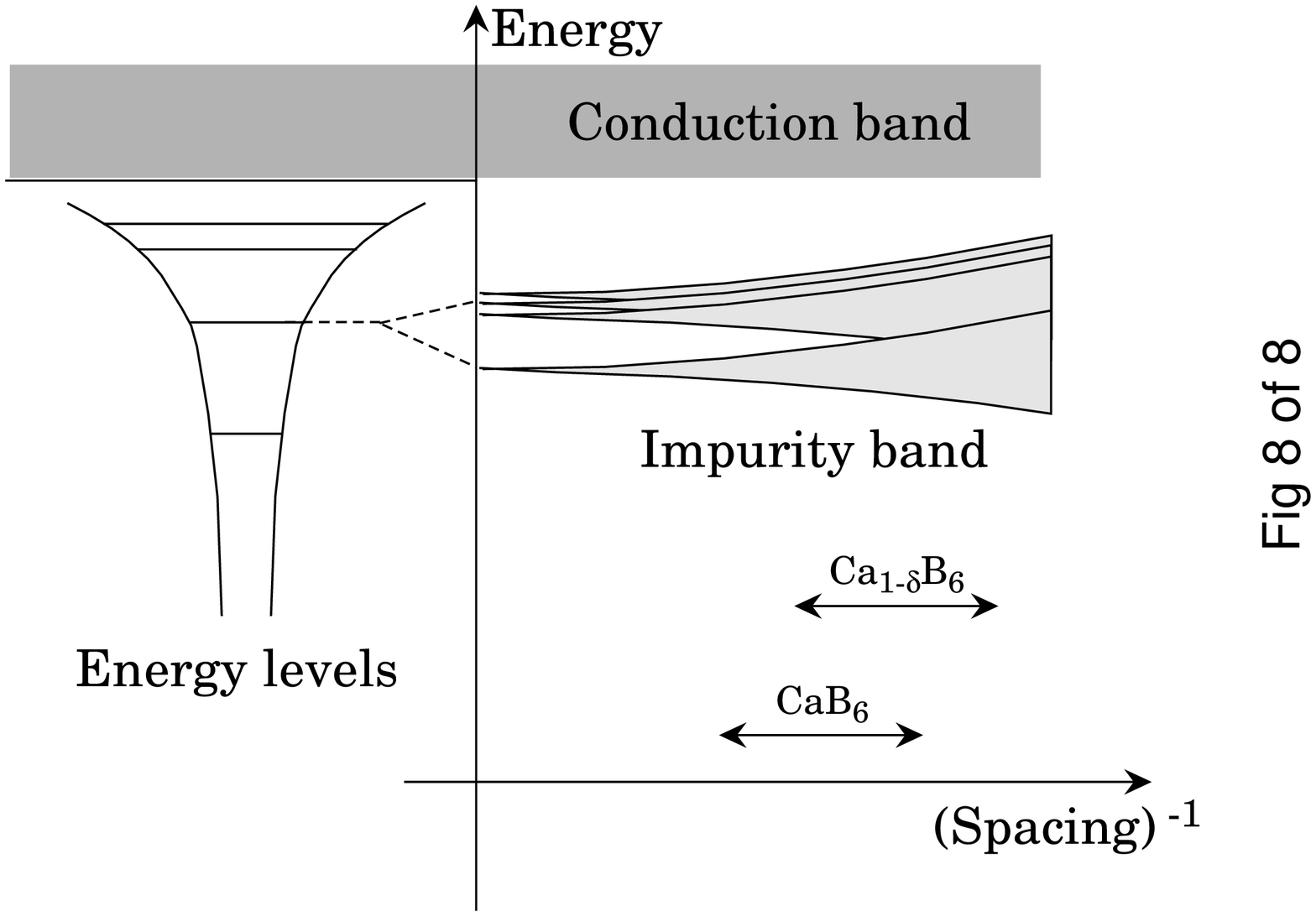}
	\vspace{1.0ex plus 0.5ex minus 0.2ex} 
           \caption{Schematic representation of the bandwidth 
           dependence on defect spacing. A higher 
           defect concentration in the vacancy-doped sample \CaMinus\, 
           decreases the spacing between donor states, resulting in 
           a higher concentration of overlapping states. The 
           non-overlapping, localized states are postulated to act as 
           resonant scattering centers for the phonons.}
           \protect\label{CaB6Bands}
	\end{center}
\end{figure}

In order to find a more plausible explanation for the experimentally 
observed crossing of $\kappa_{\rm ph}(T)$, we note that a strong reduction of the
lattice thermal conductivity of semiconductors induced by defect donor states
has been reported in a large number of publications.~\cite{Bird,Adolf,Puhl}
The intensity of the phonon scattering processes invoking such states 
depends strongly on the electronic excitation spectra induced by these 
defects. 
In particular, donor states in cubic semiconductors, leading to the 
existence of localized singlet-triplet centers just below the conduction 
band, are also expected to provide the source for a strong  
resonant phonon scattering.~\cite{Puhl} 
Physical realizations of such systems have been found in the valley-orbit-split 
ground state of the donors As, Sb, P, and Li in Ge. 
The singlet-triplet valley-orbit splitting of the 
fourfold degenerate ground state of As in Ge, for example, is 4.23~meV, 
a value which is of the same order of magnitude as the resonant energies of 
10.85 and 9.38~meV, quoted in Table~\ref{fitpar} 
for \CaPlus\, and \CaMinus, respectively.
If the density of defects in the crystal is such that the average distance 
between the donor states is of the order of the extension of the defect wave function, 
characterized by a Bohr radius of typically 40~\AA, we expect a broadening 
of the degenerate energy-eigenvalues, finally resulting in the formation of 
a band of defect states. 
The typical extension of the defect wave function should be compared with the
expected mean distance between the electrons in \CaPlus, estimated by calculating 
the third root of the inverse of the electron density derived in 
section~\ref{sec:TEP}. This leads to values of 70~{\rm \AA}\, and 100~{\rm \AA}\, for 
\CaMinus\, and \CaPlus, respectively. If the electron densities at 0~K 
derived in section~\ref{sec:TEP} are due to the existence of donor 
states in \CaPlus, we are thus close to concentrations for which a band 
is expected to form.
A schematic representation of this is shown in Fig.~\ref{CaB6Bands}. 
Because of the statistical distribution of distances between donor states, 
it seems possible to realize a situation where both, localized donor states and 
defect-band itinerant states, coexist. 
Since stoichiometric \CaPlus\, is closer to perfection than the vacancy-doped 
sample, the observation of a stronger high-frequency scattering mechanism in
\CaPlus, as mentioned above, might appear as rather surprising. 
However, as we noted above, an increasing distance between the defects due 
to a lower concentration of defects would narrow the tight-binding bands in 
\CaPlus, eventually resulting in a larger fraction of localized donor states 
acting as resonant scattering centers for the phonons. 
Hence, if this scheme is valid, the higher resonant scattering rate in \CaPlus\, 
makes sense. 

All our attempts to cast the scattering of phonons at localized donor 
states into a reasonable equation for its relaxation rate and to 
obtain a fair description of the high temperature features of 
$\kappa_{\rm ph}$ for \CaPlus\, failed.
This failure may either be due to a breakdown of the Debye model at higher 
frequencies, or may indicate that localized donor states in \CaPlus\, 
do not scatter phonons.
With respect to the first possibility it may well be that using the Debye model at the
frequencies around the resonance is not justified. 
A strong resonant scattering mechanism may, in some cases (see, for example, 
Ref.~\onlinecite{Roundy}), modify the unperturbed dispersion relations by opening
a gap in the phonon spectrum at the resonant frequency, thus turning the Debye 
scheme into a very poor approximation at the frequencies around the resonance.
This may also be the reason for the poor quality of the fit to the \CaPlus\, 
data, for which the resonant-scattering is quite significant.
Since the dispersion relations of the lattice vibrational modes in \CaPlus\, 
have not been measured yet, we cannot offer a more detailed discussion. 
In case of deviations of the acoustic branches of \CaPlus\, from a 
non-dispersive $k$ dependence at the frequencies which 
are relevant for our investigation, the functional form of 
$\tau^{-1}_{\rm res}$ would certainly change.

Concluding this section, we recall that our thermal conductivity data show a strong 
and somewhat unexpected dependence on calcium vacancy-doping at elevated 
temperatures, which we interpret as being due to a resonant scattering of the
vibrational modes. 
For a more realistic computation of the thermal transport properties at
the high end of the temperature regime covered in our investigation, 
more information concerning the phonon dispersion relations of \CaPlus\, 
is necessary, however.

\section{Summary and conclusion}
\label{sec:summary}

We have measured the thermoelectric power and the thermal conductivity of close
to stoichiometric and vacancy-doped \CaPlus\, between 5 and 300~K. 
The high negative values of the thermoelectric power $S(T)$ of both materials 
indicate a low concentration of itinerant $n$-type charge-carriers. 
We have achieved a reasonable interpretation of $S(T)$ by using a 
relaxation-time approximation of Boltzmann's equation, by considering 
a sizeable gap between valence and conduction band states, and by postulating
the existence of an additional term in the conductivity spectrum 
which may be related to the existence of a defect band in proximity to
the lower conduction band edge. 
A Debye-type relaxation-time approximation is successful in describing  
the thermal conductivity data across the entire temperature range covered 
in the experiments. The usual terms contributing to the total scattering 
rate of the vibrational states agree well with our expectations. 
A rather unusual temperature dependence of $\kappa_{\rm ph}$ is 
observed above approximately 30~K. 
A resonant type of scattering seems to influence the mean free path of the phonons
at these temperatures. 
Quite surprisingly, its limiting effect on the time period between two vibrational-mode
collisions is much stronger in the material closer to stoichiometry. 
Although postulating a resonant-type of phonon-phonon scattering
leads to a satisfactory description of our experimental 
observation, an alternative possible interpretation which 
considers the existence of localized donor states is suggested. 
Since the lattice dispersion relations of \CaPlus\, have not yet been investigated 
experimentally, we cannot offer any rigorous statements about the physical nature of
this resonant type of scattering of the vibrational modes.    

\section{Acknowledgements}
\label{sec:acknowledgements}

We thank R. Monnier and M. E. Zhitomirsky for stimulating discussions.
This work was financially supported by the Schwei\-zer\-ische 
Na\-tio\-nal\-fonds zur F\"or\-der\-ung der Wis\-sen\-schaft\-lich\-en 
For\-schung.

\newpage
    
%
%
\begin{table}[hbp]
\vskip-\lastskip
\caption{Parameters of the fitting of the $\kappa_{\rm ph}(T)$-data to 
Eqs.~\ref{Debye} and \ref{tau}.} 
\label{fitpar}
\begin{tabular}{llcc}
\multicolumn{1}{l}{Fit parameter} &
\multicolumn{1}{l}{Units} &
\multicolumn{1}{c}{\CaPlus} &
\multicolumn{1}{c}{\CaMinus} \\  
\tableline 
$l_{\rm Cas}$      &  $\mu$m               & $303\pm 4$        & $83\pm 1$       \\ 
$A$                &  10$^{-6}$            & $20.5\pm 0.1$     & $18.0\pm 0.5$   \\ 
$B$                &  10$^{-44}$ sec$^3$   & -                 & $7.7\pm 0.1$     \\ 
$C$                &  10$^{10}$ sec$^{-1}$ & $98.5\pm 0.1$     & $7.9\pm 0.1$    \\
$\hbar\omega_{\rm res}$ &  meV             & $10.847\pm 0.003$ & $9.38\pm 0.02$ 
\end{tabular}
\end{table}
%
%

\newpage

%
%
%
\begin{figure}[t]
        \begin{center}
	\leavevmode
	\epsfxsize=0.7\columnwidth \epsfbox {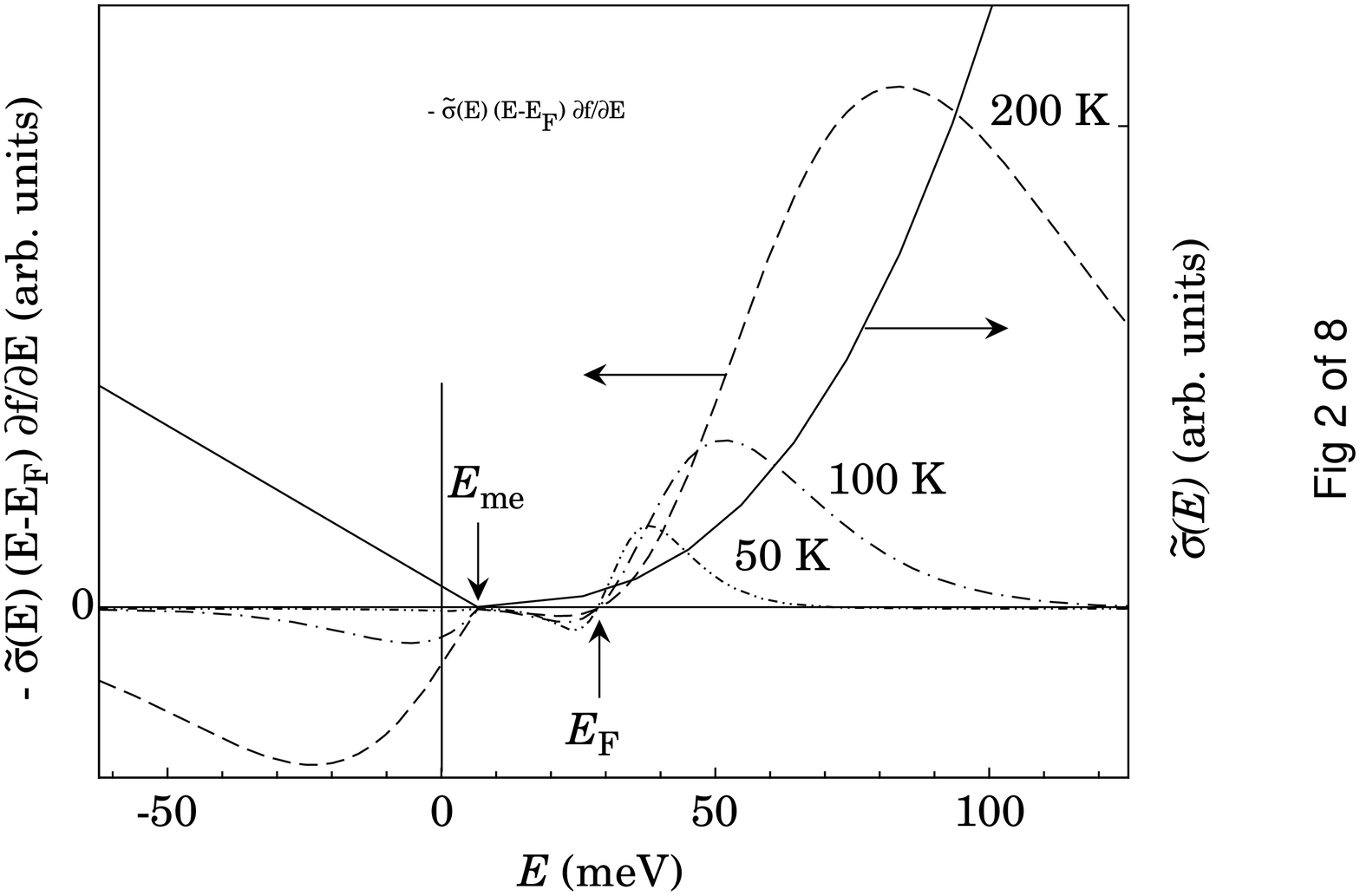}
	\vspace{1.0ex plus 0.5ex minus 0.2ex} 
           \caption{Auxiliary quantities, relevant in our analysis 
           of $S(T)$. The solid 
           line is the postulated conductivity spectrum $\tilde\sigma(E)$.
	   The various broken lines represent the value of the integrand 
           appearing in $L_{1}$ as a function of energy for 
           $T=50,100,200~{\rm K}$, as explained in the text.  
           In the inset, the integrand of $L_{1}$ is shown as a 
           function of $T$ and $E$. For a given temperature, the 
           thermoelectric power $S$ is proportional to the integral 
           taken along the curves shown on the three-dimensional 
           surface.}
           \protect\label{CaB6Motivation}
	\end{center}
\end{figure}
%
%
%
%
%
%
%
%
%
%
%
%
%
\end{document}